# Fe-vacancy Ordered Fe$_4$Se$_5$—The Insulating Parent Phase of FeSe Superconductor

**Keng-Yu Yeh[1,2,4], Yan-Rai Chen[1,2,4], Tung-Sheng Lo[1], Phillip M. Wu[1,5], Ming-Jye Wang[1,3], Kuei-Shu Chang-Liao[4], and Maw-Kuen Wu[1*]**

[1]Institute of Physics, Academia Sinica, Taipei 115, Taiwan

[2]Taiwan International Graduate Student Program, Academia Sinica, Taipei 115, Taiwan

[3]Institute of Astronomy and Astrophysics, Academia Sinica, Taipei 115, Taiwan

[4]Department of Engineering and System Science, National Tsing Hua University, Hsinchu 300, Taiwan

[5]BitSmart LLC, San Mateo CA 94403, USA

**\* Correspondence:**
Maw-Kuen Wu
mkwu@phys.sinica.edu.tw

**Abstract**

We have carried out a detailed study to investigate the existence of an insulating parent phase for FeSe superconductor. The insulating Fe$_4$Se$_5$ with $\sqrt{5} \times \sqrt{5}$ Fe-vacancy order shows a 3D-Mott variable range hopping behavior with a Verwey-like electronic correlation at around 45 K. The application of the RTA process at 450 °C results in the destruction of Fe-vacancy order and induces more electron carriers by increasing the Fe$^{3+}$ valence state. Superconductivity emerges with $T_c \sim$ 8K without changing the chemical stoichiometry of the sample after the RTA process by resulting in the addition of extra carriers in favor of superconductivity.

## 1 Introduction

The FeAs-based [1] and FeSe-based [2] superconductors are among the most investigated materials in condensed matter physics since their discovery in 2008. The observation of a wide range of superconducting transition temperatures, with the highest confirmed Cooper pair formation temperature up to 75 K in monolayer FeSe film [3] provides a unique opportunity to gain more insight into the origin of high-temperature superconductivity. The multiple-orbital nature of the Fe-based materials, combined with spin and charge degrees of freedom, results in the observation of many intriguing phenomena such as structural distortion, magnetic or orbital ordering [4], and electronic nematicity [5, 6]. There are suggestions that the orbital fluctuation may provide a new channel for realizing superconductivity [7, 8].

The parent compounds of FeAs-based materials exhibit structural transitions from a high-temperature tetragonal phase to a low-temperature orthorhombic phase, which accompanies with an antiferromagnetic (AF) order [9, 10]. Upon doping, both the orthorhombic structure and the AF phase are suppressed and superconductivity is induced. On the other hand, FeSe undergoes a tetragonal-to-orthorhombic transition at ~ 90 K [2, 11, 12]. However, no magnetic order is formed at ambient pressure [12, 13] and superconductivity below ~8 K [2] is crucially related to this orthorhombic distortion. The nematic order coexists with superconductivity but not with long-range magnetic order has led to arguments that the origin of the nematicity in FeSe is not magnetically but likely orbital-driven [14, 15].

However, recent studies show that the nematic states in the FeSe systems are far more complex [16–20]. There exist strong high-energy spin fluctuations [20] which suggest that the nematicity and magnetism may be still intimately linked. It was also found that there are many interesting features in the band structures of the nematic state. More surprises came as one applied pressure to FeSe. The application of pressure leads to the suppression of structural transition, the appearance of a magnetically ordered phase at ~1 GPa [13, 21], and $T_c$ increases to a maximum ~ 37 K [22–27] at ~6 GPa. An even more dramatic enhancement of $T_c$ was achieved on monolayer FeSe grown on $SrTiO_3$ substrate [28–31]

The above observations lead to questions that exist since the discovery of FeSe superconductor: what is the exact chemical stoichiometry of the compound, and what is the exact phase diagram for the FeSe system? Earlier studies showed that the superconducting property of FeSe is very sensitive to its stoichiometry [2, 12, 32]. The fact that higher superconducting transition temperature exists in monolayer FeSe on $SrTiO_3$ substrate suggests that the commonly accepted phase diagram, derived from assuming that FeTe is the non-superconducting parent compound of FeSe [33], is questionable. Studies have observed the trace of the superconducting feature with $T_c$ close to 40 K in samples of nano-dimensional form [34].

It has been a debate whether there exists an antiferromagnetic Mott insulating parent phase, similar to the cuprate superconductors, for FeSe superconductors [35, 36, 37]. Chen et al. first reported the existence of tetragonal β-$Fe_{1-x}Se$ with Fe vacancy orders, characterized by analytical transmission electron microscopy [38]. The authors further argued that the $Fe_4Se_5$ phase with √5 × √5 Fe-vacancy order to be the parent phase of FeSe superconductors [38]. The Fe vacancy order observed in the $Fe_4Se_5$ phase is identical to the Fe-vacancy order observed in the $A_2Fe_4Se_5$ (A=K, Tl, Rb), which has been shown to be an antiferromagnetic [35, 39-42] and is the parent phase of the superconductor $A_{2-x}Fe_{4+x}Se_5$ [43, 44]. The detailed studies of the Fe vacancy in $K_2Fe_{4+x}Se_5$ reveal that its order/disorder is directly associated with superconductivity. A recent study shows that the Fe-vacancy ordered $Fe_4Se_5$ nanowire is the non-oxide material with the Verwey-like electronic correlation [45]. It suggests that a charge-ordered state emerges below $T$ = 17 K. The question remains unanswered is whether this Fe-vacancy ordered phase is the parent compound of superconducting FeSe?

In this paper we present the results of structure, electrical transport, and magnetic measurements on the polycrystalline sample of $Fe_4Se_5$ treated by rapid-thermal-annealing (RTA) process at a proper temperature and time. After RTA treatment, the sample shows superconductivity with $T_c$ ~ 8 K without changing its chemical stoichiometry. Our findings confirm that the $Fe_4Se_5$ with Fe-vacancy order is the parent compound of FeSe superconductors.

## 2 Experimental Techniques

### 2.1 Sample Preparation

$Fe_4Se_5$ nanosheets were prepared by a chemical co-precipitation method. First, 200 mL of ethylene glycol was mixed with NaOH and $SeO_2$ powder and slowly heat up to 160 °C for mixing well.

The volume of 2.4 mL hydrazine hydrate was then added as the reducing agent. Then, at 160 °C, the Fe precursor solution was added and reacted for 12 hours in order to form $Fe_4Se_5$ nanosheets. The Fe precursor solution is made by dissolving the amount $FeCl_2$ in ethylene glycol. The reaction was done under $N_2$ gas purging to avoid the formation of oxide impurity. To clean the $Fe_4Se_5$ nanosheets, the reacted product .was dispersed in acetone with absolute ethanol, and high-speed centrifugation is applied to precipitate the nanosheets and remove the capping ligand dissolved in the above organic solvent. The nanosheets were finally dried in vacuum for 24 hours and collected. The process for the rapid thermal annealing (RTA) is: the as-grown $Fe_4Se_5$ nanosheets were heated at 450 °C for 10 minutes in a tube furnace with 1atm Ar gas inside to maintain a non-oxidation environment as a rapid thermal treatment process. After the rapid thermal treatment, an air-quenching process was taken by flowing room-temperature Ar gas through the tube. All the samples were stored in the oxygen-free glove box.

## 2.2 Analysis

The crystal structure observation of the $Fe_4Se_5$ samples was carried out by the high resolution transmission electron microscope (HRTEM, JEOL JEM-2100F) and 4-circle x-ray diffractometer with the incident beam (12.4 keV) of wavelength 0.82656 Å at beam-line BL13A and wavelength 0.61992 Å at beam-line TPS09A in NSRRC. The temperature-dependent structural information of $Fe_4Se_5$ samples was analyzed by the high resolution neutron powder diffraction (high resolution powder diffractometer Echidna with the wavelength of 2.4395 Å at ANSTO). The $Fe_4Se_5$ nanosheet powder was pressed into the pallet under 200 kg/cm$^2$ at 100 °C for 1 hour for the following measurements: To identify the stoichiometry, the energy-disperse X-ray spectrometer (EDS) setup with the SEM (JEOL JSM-7001F Field Emission Scanning Electron Microscope) was applied. To affirm the valence states of the Fe ions, the X-ray photoemission spectroscopy (XPS, VG Scientific ESCALAB 250) measurements for the samples were performed. For polycrystalline bulk samples, the resistance was measured by using the standard four-probe method with silver paste for electrical contact and the Hall measurement by a Hall-bar configuration was done by the Quantum Design Physical Properties Measurement System (PPMS, Model 6000). The magnetic property was measured by the Quantum Design Superconducting quantum interference device (SQUID, VSM).

## 3 Experimental Results and Discussions

## 3.1 Structural analysis of $Fe_4Se_5$

**Figure 1(A)** shows the X-ray diffraction (XRD) patterns of the as-grown $Fe_4Se_5$ sample at room temperature. The diffraction pattern, which exhibits superstructure peaks, is refined with a tetragonal *P4* symmetry with $\sqrt{5}\times\sqrt{5}$ Fe-vacancy order instead of the tetragonal *P4/mmm* symmetry as observed in FeSe [2]. The insets are the TEM image of as-grown $Fe_4Se_5$ nanocrystal and its TEM-SAED (selective area electron diffraction) patterns along the *c*-axis. The observation of extra diffraction points among the main diffraction points in the SAED pattern confirms the $\sqrt{5}\times\sqrt{5}$ Fe-vacancy order in the as-grown $Fe_4Se_5$ nanocrystal [35]. After RTA treatment at 450 °C, the superstructure peaks observed in XRD and TEM-SAED due to the $\sqrt{5}\times\sqrt{5}$ Fe-vacancy order disappears, as shown in **Figure 1(B)** and its inset. The refinement results, using the same *P4* symmetry, reveal that the occupations of Fe at vacancy 4d sites and the originally occupied 16i sites are almost the same, indicating the Fe vacancies becoming disordered after RTA treatment. It is noted that the XRD patterns of the RTA treatment sample show nearly ten times smaller in the intensity response. This is due to the different synchrotron beamlines we used for our measurements. However, the difference does not affect the refined results.

It is noted that the FeSe4 tetrahedron in as-grown $Fe_4Se_5$ is highly distorted due to the existence of the Fe-vacancy order. As the Fe vacancies disordered by the RTA treatment, the FeSe4 tetrahedron becomes more symmetric. The refined structure parameters are tabulated in **Supplementary Table 1 and 2**. The EDS analysis confirms that the stoichiometries of samples remain $Fe_4Se_5$ with Fe/Se ratio of 44.5: 55.5 and 45.3: 54.7 before and after the RTA treatment, respectively, as shown in **Supplementary Figure 2** in the supplementary information.

### 3.2 Temperature dependence of the resistance and the magnetic susceptibility measurement of $Fe_4Se_5$

**Figure 2(A)** shows the temperature dependence of the resistance for the as-grown polycrystalline pellet sample made of $Fe_4Se_5$ nanosheets. The inset is the magnetic susceptibility of the as-grown sample from 300 K to 2 K. The R-T results of $Fe_4Se_5$ exhibit a metal-insulator transition with the sharp rise in resistance at ~45 K. The data can be well-fitted to the three-dimensional Mott variable range hopping model (3D-MVH): $\rho(T) = \rho_0 \exp(T_0/T)^{\nu}$, where $T_0$ is the variable-range-hopping characteristic temperature, and the exponent $\nu$ is $1/(d+1)$ with $d = 3$. (The fitting results are shown in **Supplementary Figure 2**. A transition temperature $T_V$ is marked as the onset temperarture of 3D-MVH behavior). The variable-range-hopping characteristic temperature $T_0$ calculated is ~1400 K for the as-grown sample. The magnetic susceptibility of the same sample shows paramagnetic behavior as the sample cools down from 300K, and a sudden drop in susceptibility appears at about the same temperature as the resistance transition temperature ($T_V$) ~45 K. The sharp resistive rise and the diamagnetic drop are the two signatures for the Verwey transition observed in $Fe_3O_4$, which occurs at 125K. These results are also in line with those reported results in the $Fe_4Se_5$ nanowire, which was recently demonstrated to exhibit the Verwey-like electronic correlation [45].

**Figure 2(B)** shows the temperature-dependence resistance for the $Fe_4Se_5$ samples after 300 °C and 450 °C RTA process. The sample after 300 °C remains to behave like semiconductor. The sample treated at 450 and 300 °C changes to metallic and becomes superconducting below ~5 K with the onset superconductive critical temperature ($T_c$) ~7.8 K, as evidenced in the upper-left inset of **Figure 2(B)**. The lower-right inset is the magnetic susceptibility, which further demonstrates the superconducting transition with onset $T_c$ ~ 7.9 K. The evolution to superconductivity in this RTA-treated sample is similar to that reported in the $K_{2-x}Fe_{4+y}Se_5$ system, where superconductivity appears after Fe vacancies becoming disordered through high temperature annealing and rapid quenching processes [43, 44].

### 3.3 XPS and Hall measurement of $Fe_4Se_5$

In order to gain more insight into the observed Verwey-like electronic correlation, XPS at room temperature and temperature-dependent Hall measurements on the samples were performed. **Figure 3(A)** is the observed XPS results for the as-grown $Fe_4Se_5$ sample. **Figure 3(A)** is the observed XPS results for the as-grown $Fe_4Se_5$ sample. The XPS spectrum clearly reveals two peaks showing the existence of mixed-valence states of Fe. The observed two peaks, at 708.5 eV and 711.5 eV can be associated with the $Fe^{2+}$ and $Fe^{3+}$ states, respectively. The best data fitting gives the ratio between $Fe^{2+}$ to $Fe^{3+}$ close to 1:1. This result is similar to that observed in the magnetite $Fe_3O_4$.

**Figure 3(B)** displays theXPS results for samples with RTA treated at 300 °C and 450 °C. After the RTA treatment, the $Fe^{3+}$ state becomes dominant. The extracted $Fe^{3+}$ ion to total Fe atoms ratio is 58.7% for 300 °C RTA-treated and 73.2% for 450 °C RTA-treated samples, respectively, indicating a substantial increase in electron carriers in these samples. It should be noted that the sample after 300 °C still exhibits temperature dependent behavior like semiconductor. No specific difference of Se 3d peak at 54.7 eV before and after the RTA process of the $Fe_4Se_5$ sample according to the XPS result, as shown in **Supplementary Figure 3**.

It is known that tetragonal FeSe is a metal with two-band based on the first-principles electronic structure calculation, for example, by T. Xiang et al., [46]. T. Xiang et. al., also reported the electronic structure of $Fe_4Se_5$ with $\sqrt{5} \times \sqrt{5}$ Fe-vacancy order is a pair checkboard antiferromagnetic insulator. The calculation shows the Fe-vacancy ordered $Fe_4Se_5$ has a single band structure with n-type carrier dominated and a band gap ~290 meV. Berlijn et al., [47] investigated the effect of disordered Fe-vacancies on the normal-state electronic structure of the alkali-intercalated FeSe system, where the $KFe_4Se_5$ exhibits exactly the same Fe-vacancy order as that in $Fe_4Se_5$. They found that the disorder of Fe-vacancy can effectively raise the chemical potential giving enlarged electron pockets without adding carriers to the system.

It is noted that as reported by Chen et al. [38], there exists a series of $Fe_xSe_y$ compounds with x/y = 1/2, 2/3, 3/4, 4/5, and etc. We have carried out a systematic study using the co-precipitation method to successfully prepare tetragonal $Fe_{(1-x)}Se$ with stoichiometry of $Fe_3Se_4$ and $Fe_4Se_5$. Based on the XPS results, the observed $Fe^{3+}/Fe^{2+}$ ratio is 2 and 1 for tetragonal $Fe_3Se_4$ and $Fe_4Se_5$, respectively, as shown in **Supplementary Figure 4 (A)** and **Figure 3 (A)** . These data imply that $Fe_3Se_4$ would be hole-doped and $Fe_4Se_5$ be electron-doped if there are additional carriers based on the simple charge balance picture by considering $Fe_3Se_4$ to be the combination of $Fe^{(2+)}Se$ and $Fe_2^{(3+)}Se_3$, whereas $Fe_4Se_5$ is from $2(Fe^{(2+)}Se)$ and $Fe_2^{(3+)}Se_3$. Indeed, our Hall measurement results show at 300 K a hole concentration of $1.20 \times 10^{19}/cm^3$ for $Fe_3Se_4$ (**Supplementary Figure 4(B)**) and electron concentration of $-6.52 \times 10^{17}/cm^3$ for $Fe_4Se_5$.

Both of the as grown and RTA treated $Fe_4Se_5$ show a single-band behavior with n-type carrier from the Hall resistivity measurements, as shown in **Supplementary Figure 5**. The Hall coefficient of the as-grown sample at room temperature is $-9.59\ cm^3/C$, corresponding to the electron carrier concentration of $6.52 \times 10^{17}\ cm^{-3}$, and the carrier concentration decreases by about a factor of 8 at the transition temperature $T_V$, as shown in **Figure 3(C)**.

After the $Fe_4Se_5$ sample is RTA-treated at 450 °C, the $Fe^{3+}/Fe^{2+}$ ratio becomes close to 3:1, which means a large number of electrons are introduced, and subsequently induced superconductivity. Indeed, the Hall measurement results, as shown in **Figure 3(D)**, show that the carrier concentration at 300 K increases to $-3.34 \times 10^{21}/cm^3$ (Hall coefficient $-1.87 \times 10^{-3}\ cm^3/C$) for 450 °C RTA-treated sample. The electron carrier concentration is about four orders of magnitude increase comparing with the as-grown $Fe_4Se_5$. Obviously, the RTA treatment disrupts the Fe-vacancy long-range order and leads to the increase of electron carriers.

### 3.4 Neutron diffraction of $Fe_4Se_5$

It is well known that the Verwey transition in magnetite exhibits a structural transition accompanying with the sharp resistive and magnetic susceptibility changes. To examine whether such a structural change exists for the as-grown $Fe_4Se_5$, we have carried out the neutron diffraction at low temperatures.

The detailed structural information of the as-grown $Fe_4Se_5$ sample measured by neutron diffraction at different temperatures is shown in **Figure 4**. At room temperature, the neutron data, consistent with XRD results, fit well with the *P4*-tetragonal symmetry. At low temperatures, a distortion appears at temperatures below 30K. The data at 5 K, with an evident peak emerge shown in the inset of **Figure 4**, indicates a possible structural change. This result further supports that the as-grown $Fe_4Se_5$ nanosheets, similar to the results observed in $Fe_4Se_5$ nanowire, shows the Verwey-like correlation. The Verwey-like transition temperature of ~45 K in nanosheets is higher than that observed in the nanowire, which was found to be ~30 K. This shows the size dependence of $T_V$, which also noticed in Verwey transition [48-50]. Currently, we are waiting for the results of the detailed high-resolution XRD at low temperatures using a synchrotron source to determine exactly the low-temperature phase and the transition temperature.

## 4 Conclusions

We have carried out a detailed study to investigate whether there exists an insulating parent phase for FeSe superconductor. Our studies unambiguously show that: (1) the $\sqrt{5} \times \sqrt{5}$ Fe-vacancy ordered $Fe_4Se_5$ a Mott insulator with Verwey-like transition at low temperature; (2) $Fe_4Se_5$ is the parent compound of the FeSe superconductors. The application of the RTA process at 450 °C disrupts Fe-vacancy order and induces more electron carriers by increasing the $Fe^{3+}$ valence state. Superconductivity emerges with $T_c \sim 8$ K without changing the chemical stoichiometry of the sample after the RTA process. Consistent with the observations in $K_2Fe_{4+x}Se_5$, superconductivity is directly related to the disappearance of Fe-vacancy long-range order. In the $Fe_4Se_5$ case, no extra Fe doping is required as the random occupation of Fe atom in the vacancy sites, resulting in the addition of extra carriers in favor of superconductivity. More detailed evolution of superconductivity by varying the RTA temperature and time is currently underway in order to gain more insight into the exact phase diagram of the FeSe superconductors.

## 5 Author Contributions

M.-J.W., and M.-K.W. designed research. K.-Y.Y., T.-S.L., and Y.-R. C. performed research. M.-J.W. and K.-S.C.-L. contributed new reagents/analytic tools. K.-Y.Y., T.-S.L., P.M.W., Y.-R. C., K.-S.C.-L., M.-J.W., and M.-K.W. analyzed data and took part in physics discussions. K.-Y.Y., M.-J.W., T.-S.L., P.M.W. and M.-K.W. wrote the paper.

## 6 Funding

The work is supported by the Ministry of Science and Technology under Grant No. MOST108-2633-M-001-001 and Academia Sinica Thematic Research Grant No. AS-TP-106-M01.

## 7 Acknowledgments

The authors appreciate very much the help from Dr. G.T Huang for synchrotron XRD measurements, and Dr. C.P. Yen for the analysis of XPS results. We thank the technical support from NanoCore, the Core Facilities for Nanoscience and Nanotechnology at Academia Sinica in Taiwan.

## 8 Conflict of Interest

The authors declare that the research was conducted in the absence of any commercial or financial relationships that could be construed as a potential conflict of interest.

## 9 Copy rights

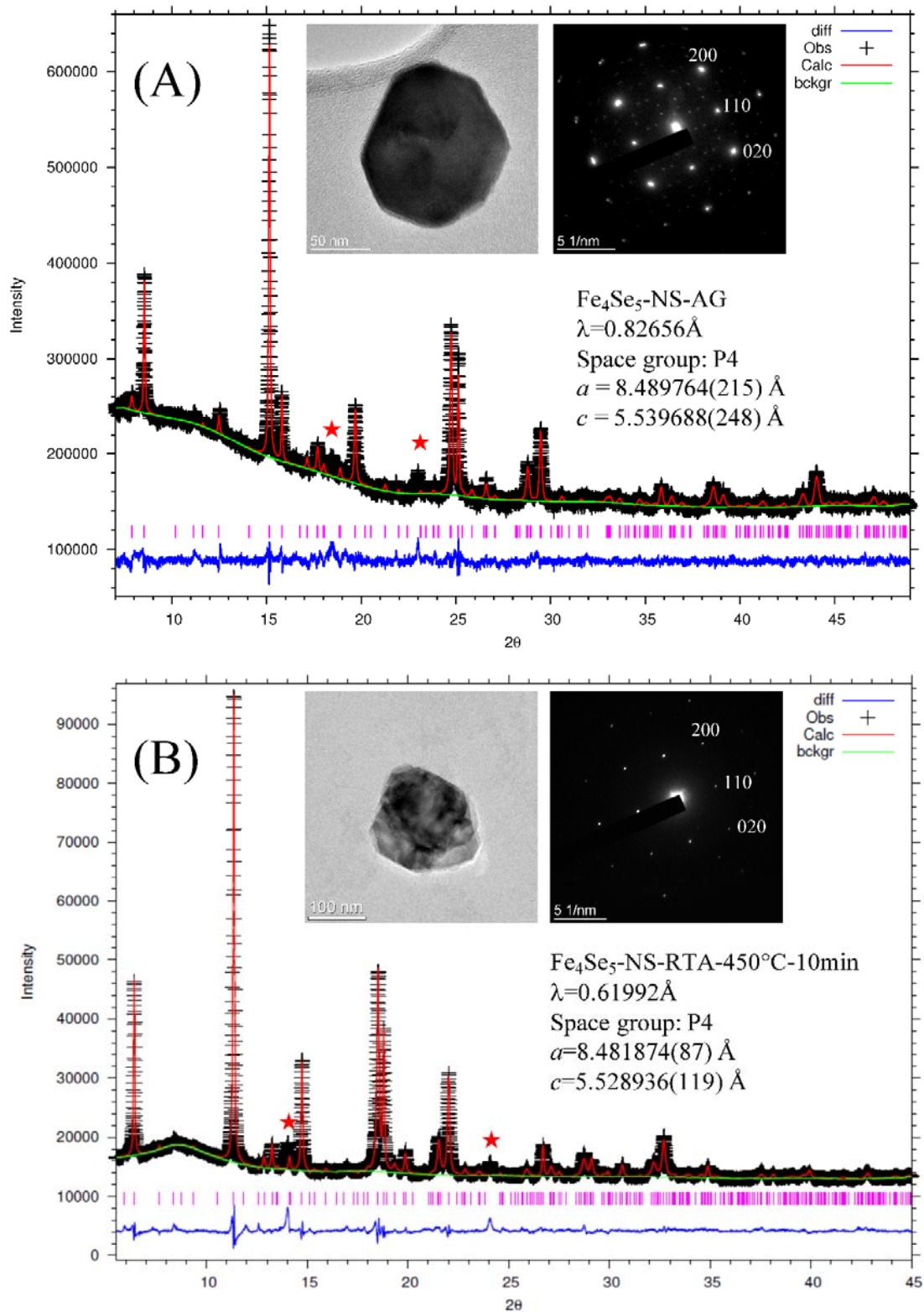

**Figure 1.** X-ray diffraction patterns and their Reitveld refinement results of **(A)** the as-grown $Fe_4Se_5$ nanocrystals and **(B)** the sample after the RTA process at 450 °C. The insets in **(A)** and **(B)** show the transmission electron microscope (TEM) images and selective area electron diffraction (SAED) patterns of the samples, respectively.

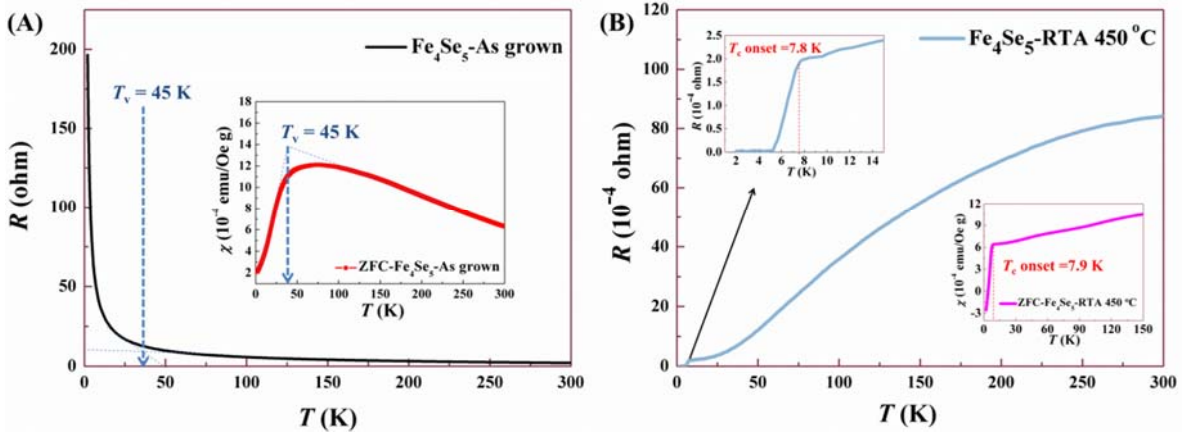

**Figure 2. (A)** Temperature-dependent resistance of the as-grown $Fe_4Se_5$ sample showing a metal-insulator transition $T_V$ with the onset of resistance rise at ~45K. The inset is the magnetic susceptibility from 300K to 2K of the same sample, which shows a large drop in susceptibility with the onset temperature also at ~45K. **(B)** Temperature-dependent resistance of the $Fe_4Se_5$ sample after 450 °C RTA process. The sample becomes metallic and shows superconducting transition with the onset $T_c$~7.8K, as highlighted in the upper-left inset. The lower-right inset is the magnetic susceptibility further demonstrates the superconducting transition with $T_c$~7.9K.

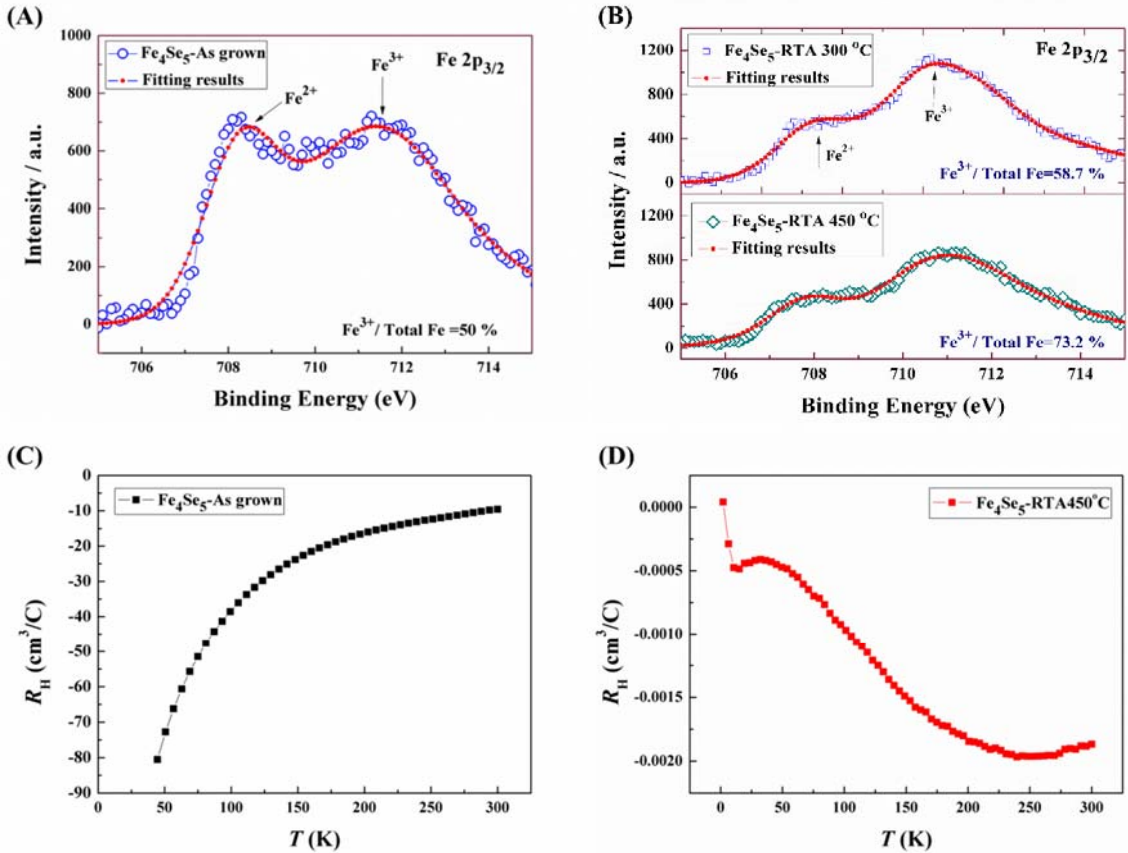

**Figure 3.** Fe$_{2p3/2}$ spectrum of XPS analysis for **(A)** the as-grown Fe$_4$Se$_5$ sample and **(B)** the 300 ºC and 450 ºC RTA-treated Fe$_4$Se$_5$ sample. The temperature dependent Hall coefficient for **(C)** the as-grown Fe$_4$Se$_5$ sample and **(D)** the 450 ºC RTA-treated Fe$_4$Se$_5$ sample.

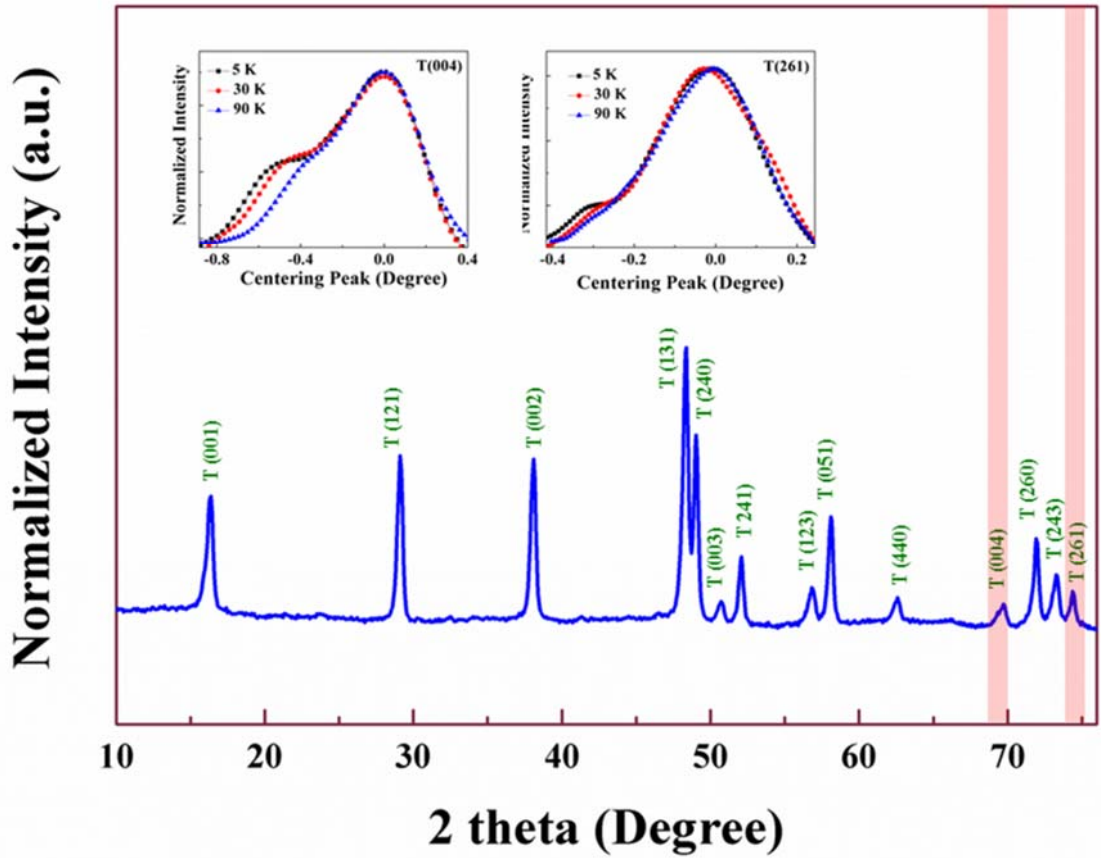

**Figure 4.** The neutron diffraction patterns of as-grown Fe$_4$Se$_5$ nanosheet at 300 K, the tetragonal *P4* symmetry is identified. The insets are the diffraction peaks of (004) and (261) at low temperatures showing the growth of additional peaks, indicating a structural change at low temperature.

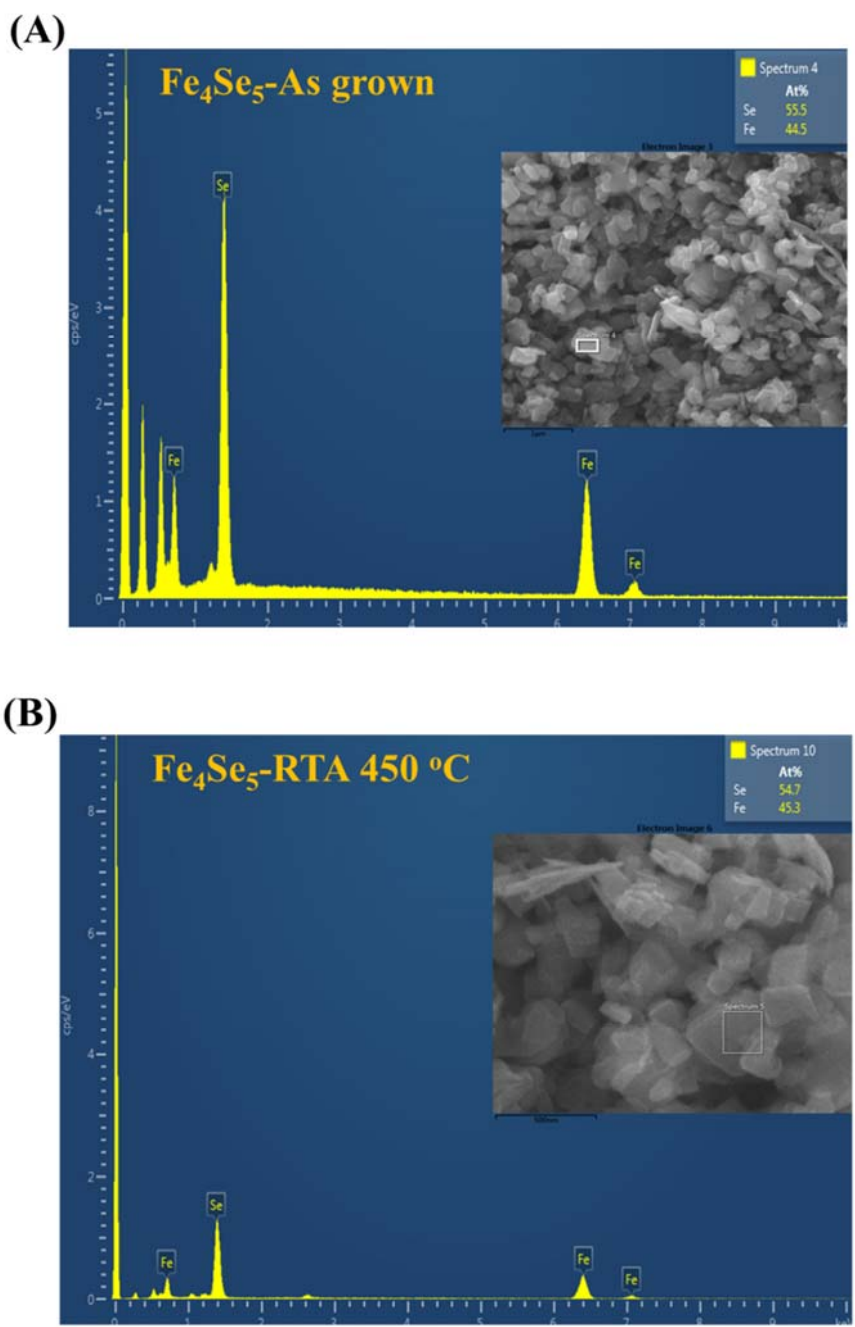

**Supplementary Figure 3.** The EDS spectrums of **(A)** the as-grown Fe$_4$Se$_5$ sample and **(B)** the RTA-treated Fe$_4$Se$_5$ sample.

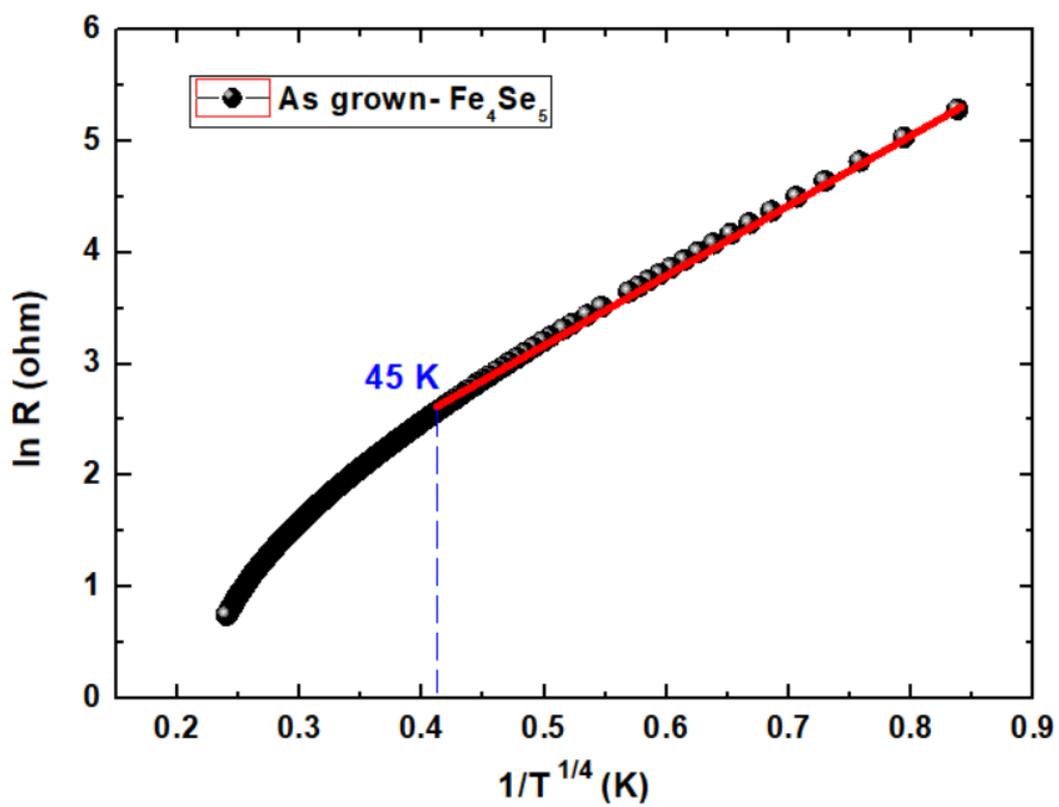

**Supplementary Figure 2.** Natural logarithmic resistance plotted against $1/T^{1/4}$ for the as grown $Fe_4Se_5$, in which the data between 2 and 45 K fit nicely by the 3D Mott variable range hopping model.

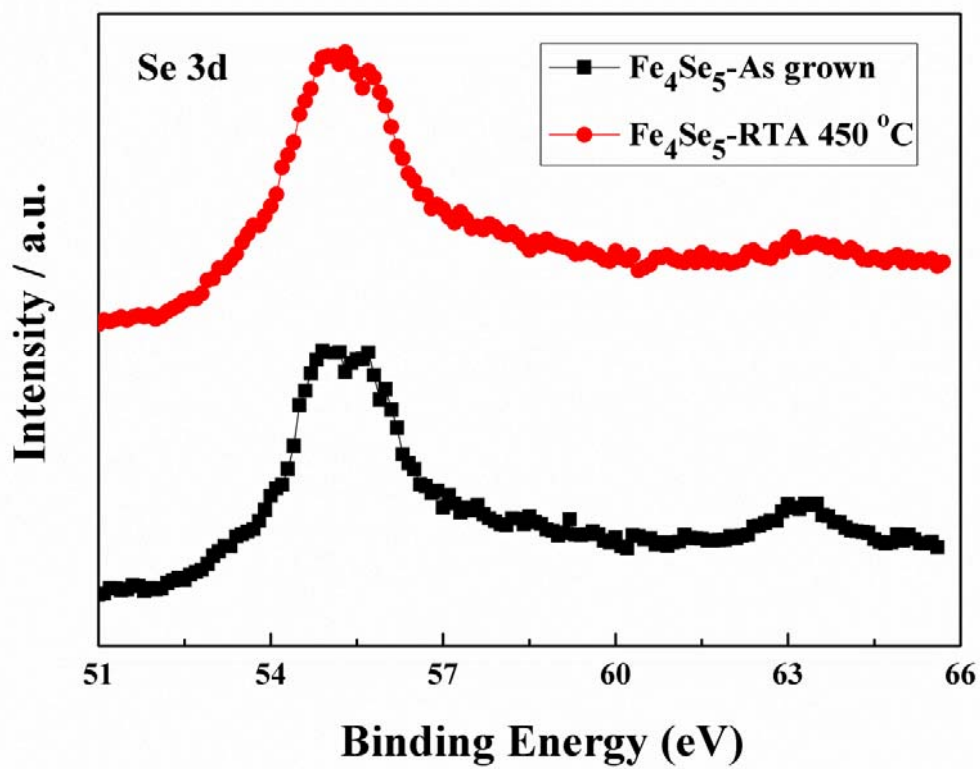

**Supplementary Figure 3**. (A) Se$_{3d}$ spectrums of XPS analysis for the as-grown Fe$_4$Se$_5$ sample and the RTA-treated Fe$_4$Se$_5$ sample.

| Space group: P4, wRp=0.0273, Rp=0.0215, $\chi^2$=74.32 $a$=8.489764(215) Å, $c$=5.539688(248) Å | | | | |
|---|---|---|---|---|
| | $x$ | $y$ | $z$ | $F$ |
| Fe-vacancy | 0.5 | 0 | 0 | 0 |
| Fe1 | 0.231613(405) | 0.084124(474) | 0.003925(2238) | 1 |
| Fe2 | 0.304264(476) | 0.427243(519) | 0.008273(2553) | 1 |
| Se1 | 0.199810(270) | 0.580714(251) | 0.259196(740) | 1 |
| Se2 | 0.079586(316) | 0.301426(291) | 0.743104(000) | 1 |
| Se3 | 0.5 | 0.5 | 0.703410(2183) | 1 |
| Se4 | 0 | 0 | 0.262407(2349) | 1 |

**Supplementary Table 1.** The structure parameters of as-grown $Fe_4Se_5$ samples as Rietveld refinement from synchrotron diffraction data.

| | x | y | z | F |
|---|---|---|---|---|
| Space group: P4 symmetry, wRp=0.0366, Rp=0.0267, $\chi^2$=8.613 $a$= 8.481874(87) Å, $c$= 5.528936(119) Å | | | | |
| Fe-vacancy | 0.5 | 0 | -0.002332(6026) | 0.806(5) |
| Fe1 | 0.200949(354) | 0.099602(275) | 0.003730(6082) | 0.776(3) |
| Fe2 | 0.301253(347) | 0.397302(259) | -0.007367(5753) | 0.847(3) |
| Se1 | 0.199458(226) | 0.597664(141) | 0.268049 (4531) | 1 |
| Se2 | 0.098462(150) | 0.299195(200) | 0.745687(4541) | 1 |
| Se3 | 0.5 | 0.5 | 0.740596(4441) | 1 |
| Se4 | 0 | 0 | 0.259767(4405) | 1 |

**Supplementary Table 2**. The structure parameters of RTA-treated $Fe_4Se_5$ samples.

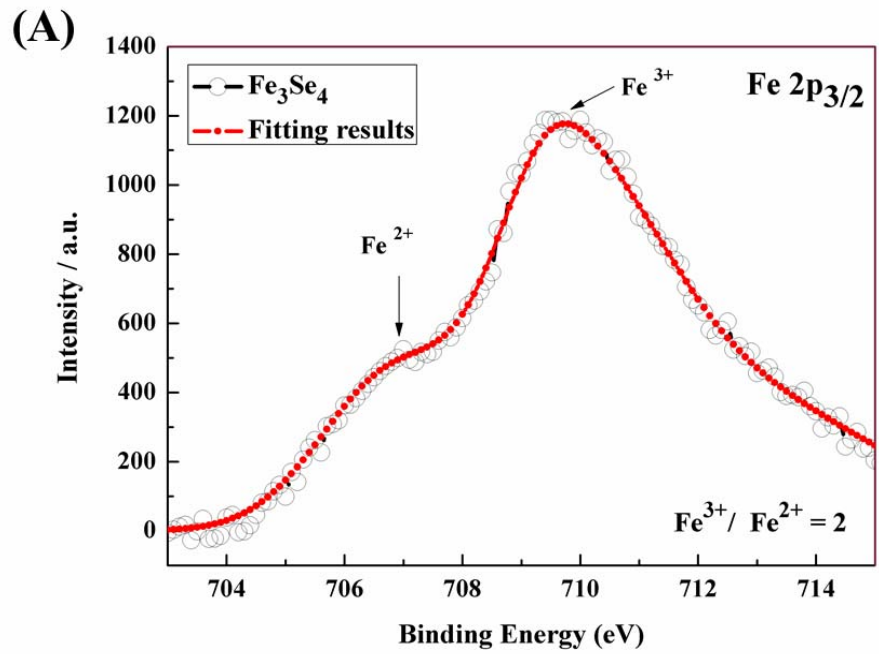

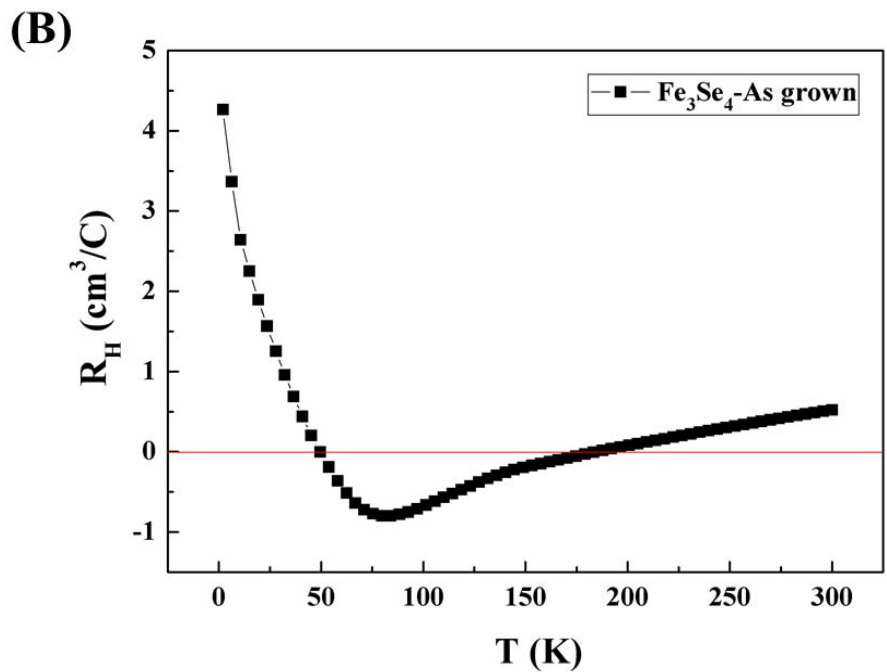

**Supplementary Figure 4**. **(A)** Fe$_{2p3/2}$ spectrum of XPS analysis and **(B)** the temperature dependent Hall coefficient for the as-grown Fe$_4$Se$_5$ sample.

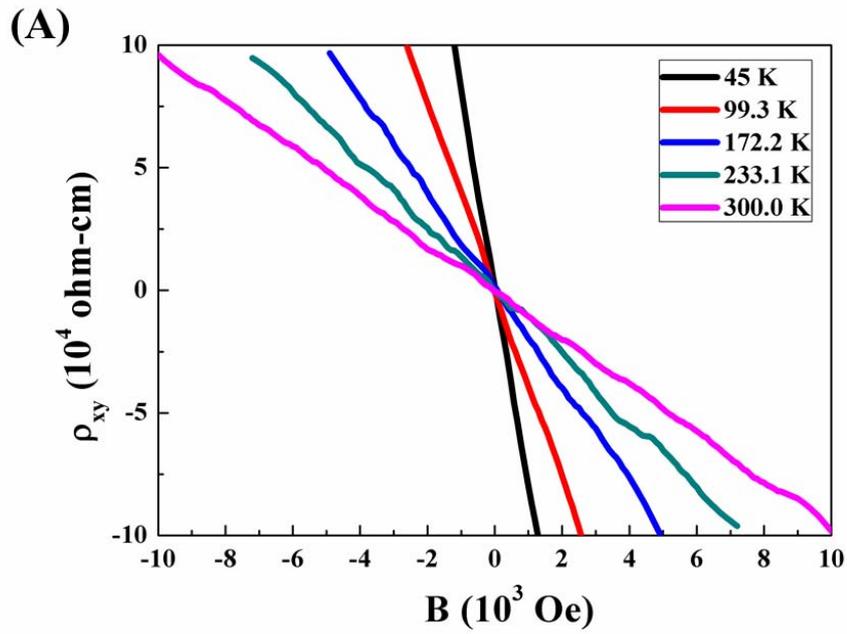

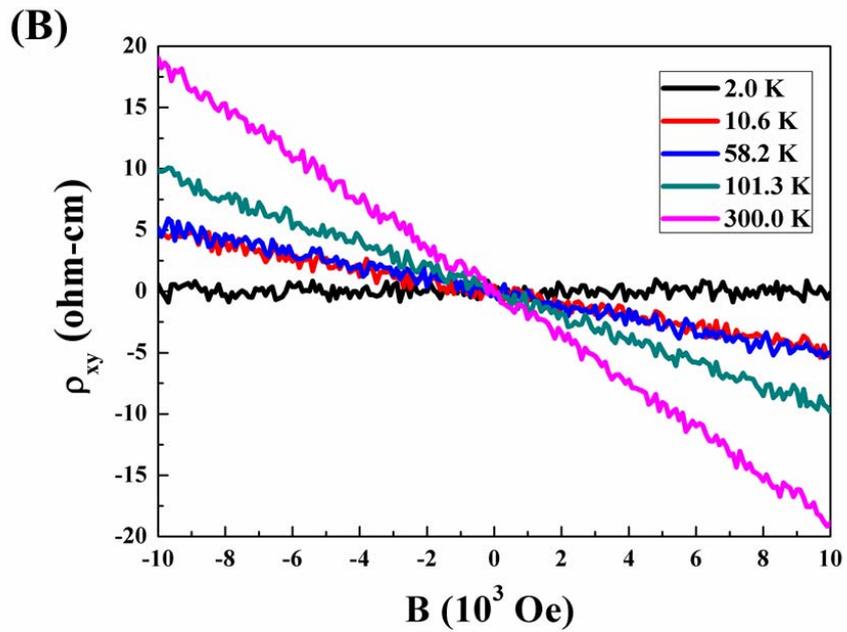

**Supplementary Figure 5**. The Hall measurement of $R_{xy}$ *v.s.* applied magnetic field of (A)the as grown $Fe_4Se_5$ nanosheet and (B)$Fe_4Se_5$ treated with RTA 450 °C under different temperature